\documentclass[doublecol]{epl2} 

\usepackage{amsmath, amssymb}

\title{Mechanisms of localization in isotope substituted dynamical Jahn--Teller systems}
\shorttitle{Mechanisms of localization of Jahn--Teller dynamics by isotope substitution}

\author{
Naoya Iwahara\inst{1,2} \and 
Tohru Sato\inst{2} \and 
Kazuyoshi Tanaka\inst{2} \and 
Liviu F. Chibotaru\inst{1}\thanks{E-mail: \email{Liviu.Chibotaru@chem.kuleuven.be}}
}
\shortauthor{N. Iwahara \etal}

\institute{                    
  \inst{1} Division of Quantum and Physical Chemistry, Katholieke Universiteit Leuven -
           Celestijnenlaan 200F, B-3001 Leuven, Belgium\\
  \inst{2} Department of Molecular Engineering, Graduate School of Engineering, Kyoto University -
           Kyoto 615-8510, Japan
}

\abstract{
The mechanisms of localization of Jahn--Teller deformations and vibronic wavefunctions
in isotope substituted dynamical Jahn--Teller systems are elucidated.
It is found that the localization in the trough is of potential type in the case of
strong vibronic coupling, while it becomes of kinetic type in the case of intermediate
and weak coupling.
It is shown that the vibronic levels in the linear $E \otimes e$-problem  remain double
degenerate upon arbitrary isotope substitution on the reasons similar to time reversal symmetry
in which the role of spin is played by orbital pseudospin.
}

\begin{document}

\maketitle

%
%
%
%
%
%

\section{Introduction}
\label{SEC:Introduction}
The Jahn--Teller (JT) effect has attracted much attention because
it is not only one of the intriguing topics in molecular physics
\cite{Bersuker1989a, Bersuker2006a}
but also crucial to the optical and magnetic properties of point defects,
electronic properties of magnetoresistive manganites,
fullerides, etc. \cite{Imada1998a, Tokura2000a, Malguth2008a, Gunnarsson1997a, Chibotaru2005a, Manini2010a}.
The properties of these materials strongly depend on the nature of JT effect,
whether it is static or dynamic.
In the study on the JT effect, the structure of the adiabatic
potential energy surfaces (APES) plays a crucial role
because it determines the distribution of the vibronic wavefunction
and hence allows to predict its localization at some distorted nuclear configuration.
Although this viewpoint has fundamental importance in the
JT problems, still other mechanisms of localization exist, e.g., those
induced by the isotope substitution
in high symmetry JT molecules and fragments.

The isotope effect in JT systems has been observed in
the spectroscopy of several molecules
\cite{Lawler1964a, Carrington1965a, Yu1993a, Bruckmeier1994a, 
Kramer1999a, Worner2007a, Melnik2011a}.
In these investigations, the isotope effect has been explained
on the basis of the favored JT distorted structures which
arise through the variation of zero-point vibrational energy
\cite{Carrington1965a, Worner2007a, Melnik2011a}.
However, no mechanisms of the isotope induced localization of JT distortions have been
established so far.
In this Letter we derive general features of the transformation
of vibronic states upon isotope substitution and find the mechanisms
of their localization.

\section{Isotope substituted $E \otimes e$ problem}
\begin{figure}[tbh]
\onefigure[height=5cm]{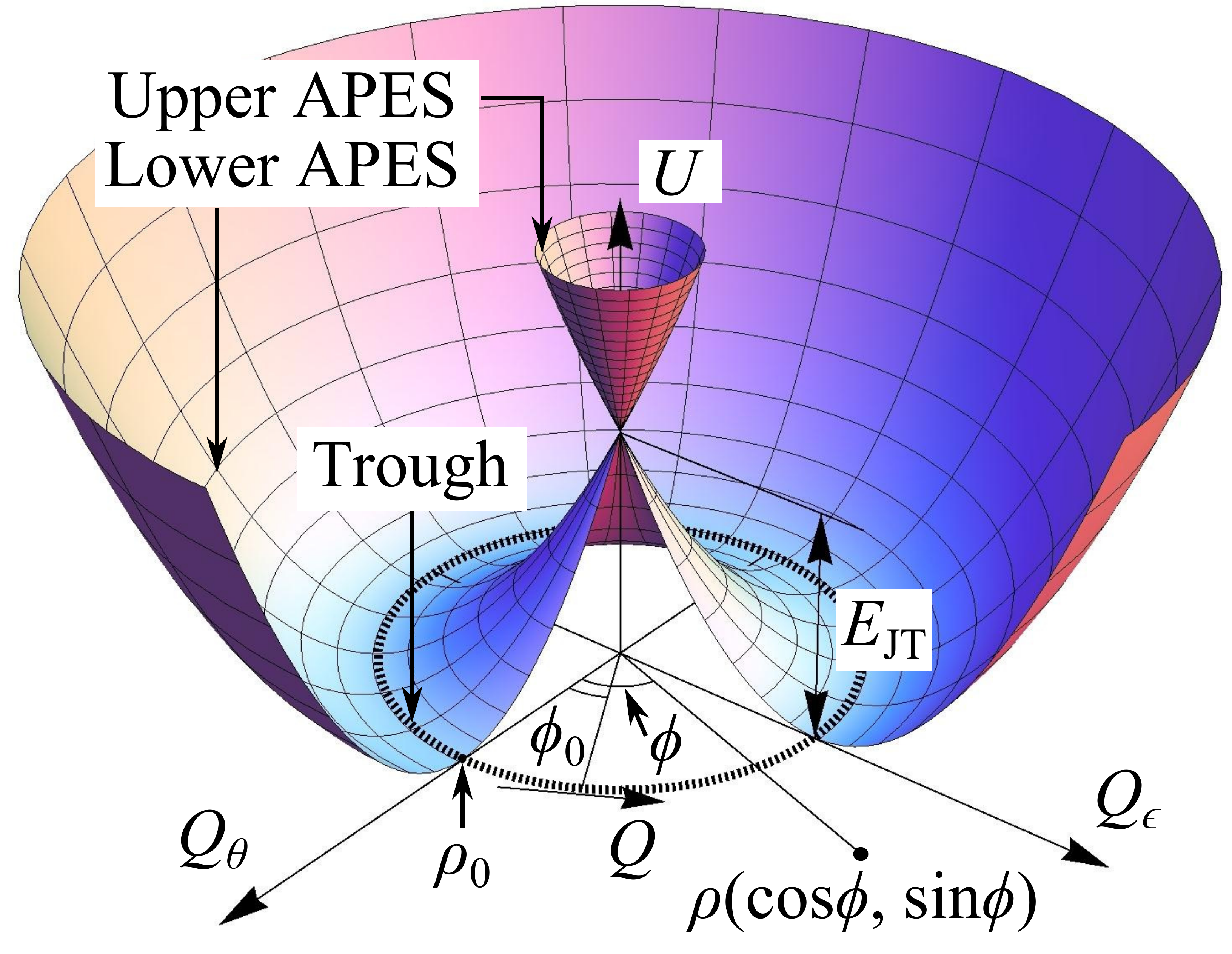}
\caption{
The APES of a linear $E \otimes e$ JT system.
The dotted circle indicates the trough.
The coordinate $Q$ tangential to the trough at $\phi_0$ corresponds to the coordinate $Q$ in fig. \ref{Fig:APES_2D}.
}
\label{Fig:APES_3D}
\end{figure}

Consider the simplest JT system with the trough in the ground APES -- the linear
$E \otimes e$ problem (fig. \ref{Fig:APES_3D})
\cite{Bersuker1989a, Bersuker2006a}. 
The linear $E \otimes e$ JT Hamiltonian is 
\begin{eqnarray}
 \hat{H}_0 &=& 
 \sum_{\mu} \left(\frac{1}{2} \hat{P}_\mu^2
 + \frac{\omega_\mu^2}{2}Q_\mu^2\right)\hat{\sigma}_0
  + V_E \left(Q_\theta \hat{\sigma}_z + Q_\epsilon \hat{\sigma}_x \right),
\nonumber\\
\label{Eq:HEeJT}
\end{eqnarray}
where $\mu$ indicates a normal mode,
$Q_\mu$ is the mass-weighted normal coordinate for the mode $\mu$,
$\hat{P}_\mu$ the conjugate momentum of $Q_\mu$, 
$\omega_\mu$ the frequency for the mode $\mu$, 
$V_E$ the vibronic coupling constant for the $e$ mode,
$\hat{\sigma}_0$ the $2 \times 2$ unit matrix,
and $\hat{\sigma}_x$ and $\hat{\sigma}_z$ the Pauli matrices.
We consider the harmonic Hamiltonian in eq. (\ref{Eq:HEeJT}) includes not only 
the $e$ mode but also other modes such as totally symmetric $a$ modes.
We also consider the system at rest, placed in the center of mass reference frame and 
neglect the coupling between the rotation and vibrations.

By isotope substitution, the mass of nuclei changes while electronic states do not.
In particular, the trough at the bottom of the lowest APES will remain unchanged (fig. \ref{Fig:APES_3D}).
In the JT Hamiltonian (\ref{Eq:HEeJT}), the mass of nuclei appears in the kinetic energy operator.
Using the Cartesian coordinates, 
the operator which expresses the isotope effect is written as
\begin{eqnarray}
 \Delta \hat{T} &=& \sum_{i,r} 
 \left(
  \frac{1}{m_i} - \frac{1}{m_i^0}
 \right) \frac{\hat{p}_{ir}^2}{2} \hat{\sigma}_0,
\label{Eq:DeltaT_Cart}
\end{eqnarray}
where $i$ indicates a nucleus in the system, 
$r (= x,y,z)$ the coordinate,
$m_i^0$ and $m_i$ the masses of nucleus $i$ and its isotope, respectively,
and $\hat{p}_{ir}$ the momentum.
The total Hamiltonian of the isotope substituted system, $\hat{H}$, is 
the sum of eqs. (\ref{Eq:HEeJT}) and (\ref{Eq:DeltaT_Cart}).

Equation (\ref{Eq:DeltaT_Cart}) is treated in two ways.
First, we transform the mass-weighted normal coordinates of the linear JT system,
$Q_\mu$, into those of the isotope substituted system, $q_\alpha$.
Compared with the linear $E \otimes e$ system, 
the symmetry of the dynamical matrix of the isotope substituted system is lowered, 
and the normal coordinate for mode $\alpha$, $q_\alpha$, becomes a linear combinations of $Q_\mu$, and vice versa: 
$Q_\mu = \sum_\alpha C_\alpha^\mu q_\alpha$.
Second, we treat eq. (\ref{Eq:DeltaT_Cart}) as a perturbation to 
$\hat{H}_0$ (\ref{Eq:HEeJT}).
The first approach is used to derive the frequencies in the trough in the strong JT coupling limit
and to perform numerical calculation. 
The second approach is used to derive analytically low-lying vibronic states at strong JT coupling.

\section{Isotope induced localization at strong JT coupling}
\begin{figure}[tbh]
\onefigure[height=4cm]{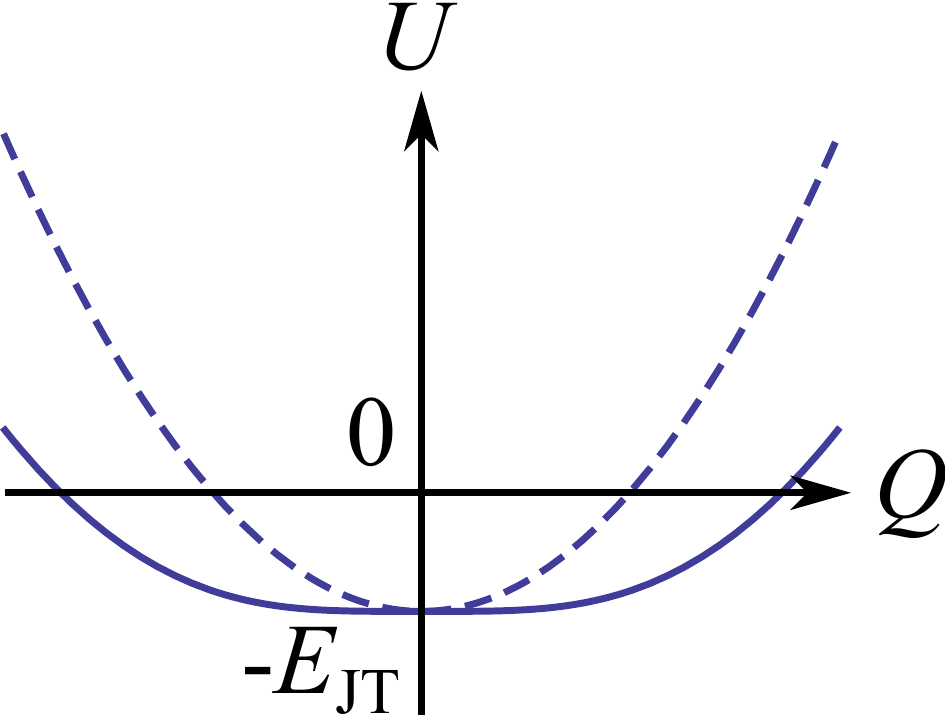}
\caption{
The cross-section of the lowest APES of a linear $E \otimes e$ JT system in the space of $(Q_\theta, Q_\epsilon)$.
The coordinate $Q$ corresponds to the tangential displacement to the trough 
at the point $\phi_0$ (fig. \ref{Fig:APES_3D}).
The dashed line is the harmonic potential for the lowest APES in the electronic state $|\psi_-(\phi_0)\rangle$,
$U = \langle\psi_-(\phi_0)|\hat{U}|\psi_-(\phi_0)\rangle$.
The solid line corresponds to the potential which 
includes the vibronic admixture of the excited electronic state $|\psi_+(\phi_0)\rangle$ by the distortion $Q$.
}
\label{Fig:APES_2D}
\end{figure}

In the strong coupling limit, 
one is justified to apply the adiabatic approximation with respect to the JT energy surfaces (fig. \ref{Fig:APES_3D}).
The corresponding adiabatic electronic states for the upper and the lower APESs, respectively, are \cite{Bersuker1989a, Bersuker2006a}:
\begin{eqnarray}
|\psi_+(\phi)\rangle &=& \cos \frac{\phi}{2}|\theta\rangle + \sin \frac{\phi}{2}|\epsilon\rangle,
\label{Eq:psi+}
\\
|\psi_-(\phi)\rangle &=& \sin \frac{\phi}{2}|\theta\rangle - \cos \frac{\phi}{2}|\epsilon\rangle,
\label{Eq:psi-}
\end{eqnarray}
where $\phi$ is the angular nuclear coordinate describing the active distortions: 
$Q_\theta = \rho \cos \phi$, $Q_\epsilon = \rho \sin \phi$ (fig. \ref{Fig:APES_3D}).
For a fixed value of parameter $\phi$ in eq. (\ref{Eq:psi-}), $\phi = \phi_0$, 
the potential energy corresponding to the ground state electronic function $|\psi_-(\phi_0)\rangle$ 
will have a minimum at $Q_\theta = \rho_0 \cos \phi_0$, $Q_\epsilon = \rho_0 \sin \phi_0$, where
$\rho_0$ is the radius of the trough (fig. \ref{Fig:APES_3D}), and zero value of other nuclear coordinates.
This situation is depicted in fig. \ref{Fig:APES_2D} by dashed line. 
%
%
Let us consider infinitesimal distortions from this minimum point, $\delta Q_\mu$. 
In terms of these infinitesimal distortions the potential energy operator becomes
\begin{eqnarray}
\hat{U} &=& 
 \begin{pmatrix}
  3E_{\rm JT} & 0 \\
  0 & -E_{\rm JT} \\
 \end{pmatrix}
 + \sum_\mu \frac{\omega_\mu^2}{2} \delta Q_\mu^2 \hat{\sigma}_0
\nonumber \\
&+& 
2V_E (\delta Q_\theta \cos \phi_0 + \delta Q_\epsilon \sin \phi_0)
 \begin{pmatrix}
  1 & 0 \\
  0 & 0 \\
 \end{pmatrix}
\nonumber \\
&+& V_E \left(\delta Q_\theta \sin \phi_0 - \delta Q_\epsilon \cos \phi_0\right)\hat{\sigma}_x,
\label{Eq:U_strongJT}
\end{eqnarray}
where $E_{\rm JT} = V_E^2/(2\omega_E^2)$ is the JT stabilization energy, 
and $\mu$ indicates all modes 
which appear in the harmonic Hamiltonian in eq. (\ref{Eq:HEeJT}).
Note that the basis in eq. (\ref{Eq:U_strongJT}), 
$\{|\psi_+(\phi_0)\rangle, |\psi_-(\phi_0)\rangle\}$, is different from that in eq. (\ref{Eq:HEeJT}).
Because of the off-diagonal term of $\hat{U}$, 
the frequency corresponding to the coordinate $Q = \delta Q_\theta \sin \phi_0 - \delta Q_\epsilon \cos \phi_0$
will reduce to zero (fig. \ref{Fig:APES_2D}) which recovers the trough at the lowest APES.
In the harmonic approximation, it is sufficient to include this off-diagonal term in the second order of perturbation theory.
The total harmonic potential energy at the point $\phi_0$ has the form
\begin{eqnarray}
U &=& 
\langle \psi_-(\phi_0)|\hat{U}|\psi_-(\phi_0)\rangle 
\nonumber\\
&+& \frac{\left|\langle \psi_+(\phi_0)|\hat{U}|\psi_-(\phi_0)\rangle\right|^2}{-4E_{\rm JT}}
\\
&=& 
\sum_\mu \frac{\omega_\mu^2}{2} \delta Q_\mu^2 -
\frac{\omega_E^2}{2} (\delta Q_\theta \sin \phi_0 - \delta Q_\epsilon \cos \phi_0)^2.
\label{Eq:U_0}
\end{eqnarray}
Now let us consider the isotope effect. 
The isotope effect is included in $U$ by substituting 
$\delta Q_\mu = \sum_\alpha C_\alpha^\mu \delta q_\alpha$ into eq. (\ref{Eq:U_0}).
\begin{eqnarray}
 U &=& 
  \frac{1}{2}\sum_{\alpha \beta} \delta q_\alpha \delta q_\beta 
  \left[
  \tilde{\omega}_\alpha^2 \delta_{\alpha \beta} -
  \omega_E^2 \left(C^\theta_\alpha \sin \phi_0 - C^\epsilon_\alpha \cos \phi_0\right)
\right.
\nonumber\\
 &\times&
\left.
  \left(C^\theta_\beta \sin \phi_0 - C^\epsilon_\beta \cos \phi_0\right)
  \right],
\label{Eq:U_iso}
\end{eqnarray}
where $\tilde{\omega}_\alpha$ is the vibrational frequency for mode $\alpha$ of the isotope substituted system
in the absence of JT coupling ($V_E=0$). 
Diagonalizing the dynamical matrix whose element is the term in the square bracket 
in eq. (\ref{Eq:U_iso}), we obtain the frequencies, $\Omega_l$, and the normal modes, $u^{(l)}_\alpha$, in the trough.
This eigenvalue problem is written as 
\begin{eqnarray}
  \sum_\beta 
  \omega_E^2 \left(C^\theta_\alpha \sin \phi_0 - C^\epsilon_\alpha \cos \phi_0\right)
  \left(C^\theta_\beta \sin \phi_0 - C^\epsilon_\beta \cos \phi_0\right) 
\nonumber\\
\times u_{\beta}^{(l)} 
= \left(\tilde{\omega}_\alpha^2 - \Omega_l^2\right) u_{\alpha}^{(l)}.
\label{Eq:eigprob1}
\end{eqnarray}
If $\sum_\beta (C^\theta_\beta \sin \phi_0 - C^\epsilon_\beta \cos \phi_0) u_{\beta}^{(l)} = 0$,
$u_\alpha^{(l)}$ becomes trivial, i.e., $u_\alpha^{(l)} = 0$ for all $\alpha$.
Therefore, $\sum_\beta (C^\theta_\beta \sin \phi_0 - C^\epsilon_\beta \cos \phi_0) u_{\beta}^{(l)} \ne 0$ 
holds for the normal vibrational mode $u_\alpha^{(l)}$.
Multiplying $(C^\theta_\alpha \sin \phi_0 - C^\epsilon_\alpha \cos \phi_0)/(\tilde{\omega}_\alpha^2 - \Omega_l^2)$
and the both sides of eq. (\ref{Eq:eigprob1}), and summing over $\alpha$,
\begin{eqnarray}
  \left[
  \sum_\alpha \frac{\omega_E^2 \left(C^\theta_\alpha \sin \phi_0 - C^\epsilon_\alpha \cos \phi_0\right)^2}
  {\tilde{\omega}_\alpha^2 - \Omega_l^2}
  - 1
  \right]
\nonumber\\
  \times
  \sum_\beta 
  \left(C^\theta_\beta \sin \phi_0 - C^\epsilon_\beta \cos \phi_0\right) u_{\beta}^{(l)} 
= 0.
\label{Eq:eigprob2}
\end{eqnarray}
The term in the square bracket in eq. (\ref{Eq:eigprob2}) is zero, which 
is the equation for the frequencies in the trough: 
\begin{equation}
 \sum_\alpha
 \frac{\omega_E^2 
  \left(C^\theta_\alpha\sin \phi - C^\epsilon_\alpha \cos \phi \right)^2}
  {\tilde{\omega}_\alpha^2 - \Omega^2}
  =1.
\label{Eq:freq}
\end{equation}
It results from this equation that, in contrast to the linear $E \otimes e$ JT system, 
the isotope substituted system does not have a zero-frequency mode.
Furthermore, we can see that the frequencies generally depend on $\phi$ because the numerators of the left hand 
sides of eq. (\ref{Eq:freq}) are functions of $\phi$.
Therefore, for the motion along the trough, warping of the potential energy surface
is induced through the zero-point energy, 
\begin{eqnarray}
E^0_{\rm vib}(\phi) &=& \frac{1}{2} \sum_l \hslash \Omega_l(\phi).
\label{Eq:E0vib}
\end{eqnarray}

In order to derive analytically the low-lying vibronic states at strong JT coupling,
we choose a triangular 
molecule X$_3$ with $D_{3h}$ symmetry as an $E \otimes e$ JT system (X = H, D, Li, etc.).
The irreducible representations of the normal modes of X$_3$ are $a_1'$ and $e'$.
Replacing one of the atoms X with its isotope Y, we obtain an isotopomer with $C_{2v}$ symmetry. 
We treat eq. (\ref{Eq:DeltaT_Cart}) as a perturbation with $C_{2v}$ symmetry to eq. (\ref{Eq:HEeJT}).
\begin{eqnarray}
\hat{H} &=& \hat{H}_0 
 + \frac{\lambda}{2} \left(\hat{P}_A^2 + \hat{P}_\theta^2 + \hat{P}_\epsilon^2 + 2\hat{P}_A \hat{P}_\theta \right) \hat{\sigma}_0,
\label{Eq:HJT_iso}
\end{eqnarray}
where $\hat{H}_0$ is the sum of the linear $E \otimes e$ JT Hamiltonian
and the Hamiltonian for the harmonic oscillator of the $a_1'$ mode, and
$\lambda = -\delta/[3(1 + \delta)]$, $\delta = (m-m^0)/m^0$.
The perturbation (\ref{Eq:DeltaT_Cart}) is rewritten using the momenta of X$_3$, $\hat{P}_\mu$.
The cross-term between $\hat{P}_A$ and $\hat{P}_\theta$ in eq. (\ref{Eq:HJT_iso}) appears because the symmetry of the perturbation is $C_{2v}$.
In eq. (\ref{Eq:HJT_iso}), we used $A$, $\theta$, and $\epsilon$ instead of $A_1'$, $E'\theta$, and $E'\epsilon$,
respectively, for simplicity of notation.

When $\lambda$ is small and $V_E$ is large, 
the Hamiltonian for the pseudorotation is derived as follows.
In the derivation, we omit constant terms and neglect terms of $O(\lambda^3)$ and $O(1/V_E^3)$.
First, diagonalizing the vibronic coupling term from $\hat{H}_0$ in eq. (\ref{Eq:HJT_iso}), 
and applying the adiabatic approximation, 
we obtain the Hamiltonian for the lower APES, $\hat{H}_{\rm ad}$ \cite{Bersuker1989a, Bersuker2006a}.
\begin{eqnarray}
 \hat{H}_{\rm ad} 
 &=& \frac{1}{2}\left[\left(1+\lambda\right)\hat{P}_A^2 + \omega_A^2 Q_A^2 \right]
\nonumber\\
 &+& \frac{1}{2}\left[\left(1+\lambda\right)\left(\hat{P}_\rho^2 + \frac{1}{\rho_0^2}\hat{P}_\phi^2\right) + \omega_E^2 \bar{\rho}^2\right]
\nonumber\\
 &+& \lambda \hat{P}_A \left[\cos \phi \hat{P}_\rho 
 - \frac{1}{2\rho_0}\left(\sin \phi \hat{P}_\phi + \hat{P}_\phi \sin \phi\right)\right],
\nonumber\\
\label{Eq:Had}
\end{eqnarray}
where $\hat{P}_\rho = -i\hslash \partial/\partial \rho - i\hslash/(2\rho)$,
$\hat{P}_\phi = -i\hslash \partial/\partial \phi$, 
$\rho_0 = V_E/\omega_E^2$, and
$\bar{\rho}$ is the displacement from $\rho_0$, $\bar{\rho} = \rho - \rho_0$.
Then, we perform the unitary transformation of the Hamiltonian to remove from eq. (\ref{Eq:Had})
the third term linear in $\lambda$, $e^{-i\hat{S}}\hat{H}_{\rm ad} e^{i\hat{S}}$.
Finally, averaging the transformed Hamiltonian by the ground vibrational state 
for the radial and the totally symmetric modes, 
the Hamiltonian, $\hat{H}_\phi$, describing the motion along the trough coordinate,
is obtained as
\begin{eqnarray}
 \hat{H}_{\phi} &=& \frac{1}{2\bar{\rho}_0^2}\hat{P}_\phi \left(1-\bar{\lambda}^2 \sin^2\phi\right)\hat{P}_\phi
  -\bar{\lambda}^2 \left( 
   \kappa_1 + \frac{\kappa_2}{\bar{\rho}_0^2}
  \right) \cos 2\phi,
\nonumber
\\
\label{Eq:H_rot}
\end{eqnarray}
where
$\bar{\rho}_0 = \rho_0/\sqrt{1+\lambda}$, 
$\bar{\lambda} = \lambda/(1+\lambda)$, 
$\kappa_1 = \hslash \sqrt{1+\lambda} \omega_A \omega_E / [8(\omega_A + \omega_E)]$, and
$\kappa_2 = \hslash^2 (3\omega_A^2 + 5 \omega_A \omega_E + 3 \omega_E^2)/[16(\omega_A + \omega_E)^2]$.
Eq. (\ref{Eq:H_rot}) will be valid when the gap of the rotational levels is several times smaller than the 
gap of the vibrational levels, $\hslash \omega_E$,
i.e., $\lambda^2 \kappa_1 \ll \hslash \omega_E \ll E_{\rm JT}$. 
Although there are no barriers in the trough of APES entering the vibronic Hamiltonian (\ref{Eq:HJT_iso}),
we obtain in eq. (\ref{Eq:H_rot}) a warped potential.
This warping is a consequence of the mixing of the vibrational states induced by the 
change of the mass in the kinetic energy,
leading to $\phi$-dependent vibrational frequencies in the trough
$\Omega_l(\phi)$ (eq. \ref{Eq:freq}) and to $E^0_{\rm vib}(\phi)$ (\ref{Eq:E0vib}).
The potential term and the kinetic term of $\hat{H}_\phi$ have two minima, which is consistent with 
the invariance of $E^0_{\rm vib}(\phi)$ and eq. (\ref{Eq:freq}) with respect to the rotation 
by $\pi$ ($\phi \rightarrow \phi + \pi$).
 
The eigenstates $\psi(\phi)$ of eq. (\ref{Eq:H_rot}) have to satisfy the boundary condition 
$\psi(\phi + 2\pi) = -\psi(\phi)$ 
to fulfill the single-valuedness of the total vibronic wavefunction. 
This sign change comes from the Berry's phase of the electronic state
of the JT system \cite{Bersuker1989a, Bersuker2006a}.
In addition, each level of $\hat{H}_{\phi}$ is obtained doubly degenerate.
We discuss the origin of this degeneracy later.

In the case of linear $E \otimes e$ problem $(\lambda = 0)$, 
the Hamiltonian (\ref{Eq:H_rot}) describes the rotation of JT deformation in the 
trough, with eigenenergies and eigenstates
$\hslash^2 j^2/2\bar{\rho}_0^2$ and $e^{i j \phi}$, respectively, where $j$ is a half-integer 
\cite{Bersuker1989a}.
These eigenstates are completely delocalized in the trough (fig. \ref{Fig:APES_3D}).
For small finite $\lambda$, $e^{i j \phi}$ mixes with $e^{i(j \pm 2)\phi}$. 
Consequently, 
the density corresponding to these vibronic eigenstates depends on $\phi$:
\begin{eqnarray}
 \left|\psi_j(\phi)\right|^2 = \frac{1}{2\pi}
 \left[1 - 
  \frac{\bar{\rho}_0^2\left(\Delta_{\rm kin} + \Delta_{\rm pot}\right)}
  {\hslash^2(j^2-1)(1-\bar{\lambda}^2/2)}\cos2\phi
 \right].
\label{Eq:psi2_rot}
\end{eqnarray}
Here, 
$\Delta_{\rm kin} = \bar{\lambda}^2 \hslash^2 j^2/4\bar{\rho}_0^2$, and 
$\Delta_{\rm pot} = \bar{\lambda}^2 (\kappa_1 + \kappa_2/\bar{\rho}_0^2)$.
$\Delta_{\rm kin}$ and $\Delta_{\rm pot}$ are the amplitudes of
the variation of the kinetic and potential terms in the trough 
(both proportional to $\cos 2\phi$).
The densities of the ground $(|j| = 1/2)$ and first excited $(|j|=3/2)$ states
are concentrated at $\phi = 0$, $\pi$ and $\phi=\pm \pi/2$, respectively.

\section{Kinetic vs potential localization}
\begin{figure}[tbh]
\onefigure{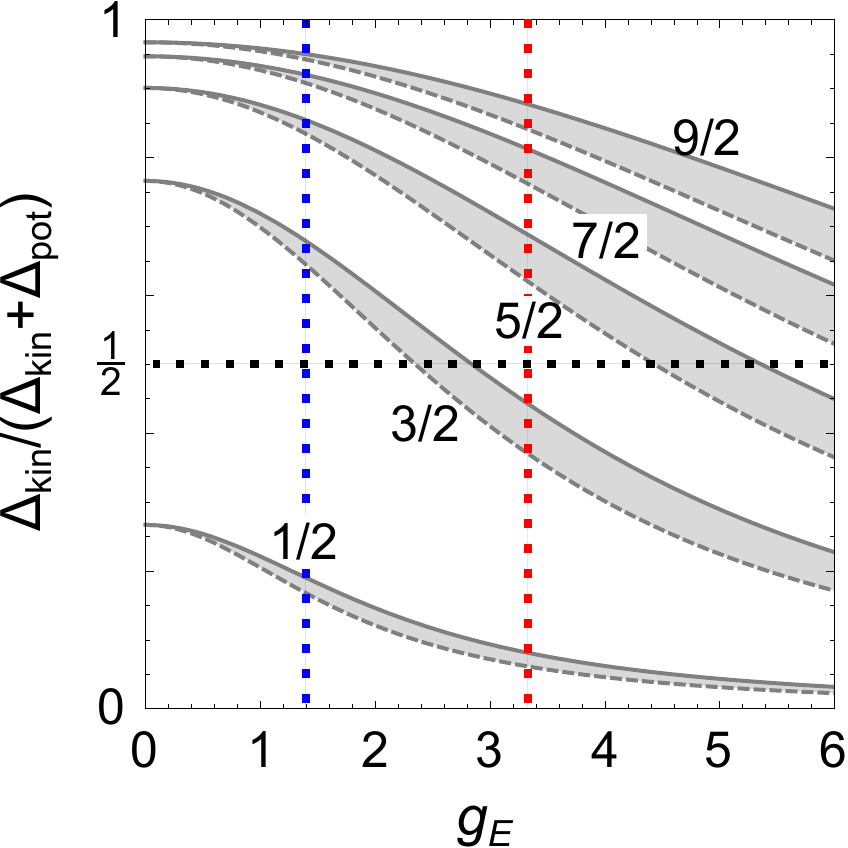}
\caption{
The kinetic contributions to the localization, $\Delta_{\rm kin}/(\Delta_{\rm kin} + \Delta_{\rm pot})$,
for the states with $j=1/2, 3/2, \cdots, 9/2$.
The grey dashed and solid lines correspond to the replacements of H $\rightarrow$ T and T $\rightarrow$ H, respectively.
When $\Delta_{\rm kin}/(\Delta_{\rm kin} + \Delta_{\rm pot}) > 1/2$, 
the kinetic term is more important than the potential term, and vice versa.
The horizontal dotted line indicates the boundary $\Delta_{\rm kin} = \Delta_{\rm pot}$,
the vertical blue dotted line ($g_E^2/2=1$) separates the weak and intermediate coupling regions, 
and the vertical red dotted line ($g_E^2/2=5$) separates the intermediate and strong coupling regions.
}
\label{Fig:kin}
\end{figure}

The localization of vibronic wavefunction along the trough is determined by the amplitudes $\Delta_{\rm kin}$
and $\Delta_{\rm pot}$.
Although both $\Delta_{\rm kin}$ and $\Delta_{\rm pot}$ appears from the change of mass, 
they have different physical meanings.
As described above, $\Delta_{\rm pot}$ is essentially the same as $E^0_{\rm vib}(\phi)$ (eq. (\ref{Eq:E0vib})).
On the other hand, $\Delta_{\rm kin}$ is characterized by the rotational quantum number $j$ 
(see the formula after eq. (\ref{Eq:psi2_rot})).
In order to assess the relative importance of the two mechanisms into the localization, 
it is interesting to analyze the ratio of the kinetic amplitude to the potential one.
The relation between the contribution of the kinetic term, $\Delta_{\rm kin}/(\Delta_{\rm kin} + \Delta_{\rm pot})$,
and the dimensionless vibronic coupling constant $g_E (= V_E/\sqrt{\hslash \omega_E^3})$ 
is shown in fig. \ref{Fig:kin}.
We put $\omega_A = \omega_E$ for simplicity.
Since $\Delta_{\rm kin}$ and $\Delta_{\rm pot}$ depend on $\delta$, we show both the cases of the largest $\delta (=2)$ 
and the smallest $\delta (=-2/3)$, which correspond to the replacement of hydrogen atom (H) by 
tritium atom (T) and the opposite, respectively.
The contribution of the kinetic term decreases as $g_E$ increases, and 
in the strong coupling limit ($g_E \rightarrow \infty$) the kinetic term vanishes ($\Delta_{\rm kin} \rightarrow 0$).
In the weak coupling limit ($g_E \rightarrow 0$), 
$\Delta_{\rm kin}/(\Delta_{\rm kin} + \Delta_{\rm pot}) \rightarrow j^2/(j^2+4\kappa_2/\hslash^2)$.
As fig. \ref{Fig:kin} shows, in the ground vibronic state {($|j|=1/2$)} the contribution of the potential term is 
dominant in the whole range of $g_E$ (localization of potential type).
On the other hand, in the excited states {($|j|=3/2,5/2,\cdots$)} the mechanism of localization depends on $g_E$.
When $g_E$ is smaller than a certain value, the contribution of the kinetic term exceeds that of the
potential term (the mechanism becomes of kinetic type).
To conclude, in the strong coupling limit the mechanism of the localization is of potential type,
while in the case of the weak $(g_E^2/2 \le 1)$ or intermediate coupling $(1 \le g_E^2/2 \le 5)$, 
the localization is of kinetic type.

One should mention the role of the Berry's phase 
in the observed localization.
In a linear $E \otimes e$ JT system, 
if the Berry's phase is ignored, $\psi(\phi + 2\pi) = \psi(\phi)$, 
the ground and first excited vibronic states would correspond to $j=0$ 
and $|j|=1$, respectively \cite{Ham1987a}.
The density distribution along the trough in the ground state of the isotopomer 
is obtained from eq. (\ref{Eq:psi2_rot}) by substituting $j=0$.
Since $\Delta_{\rm kin}=0$ for $j=0$, the kinetic term does not 
contribute to the localization.
Furthermore, $|1/(j^2-1)|$ in the right hand sides of eq. (\ref{Eq:psi2_rot})
with $|j|=1/2$ is larger than that with $j=0$.
Therefore, the Berry's phase enhances the localization of the ground vibronic wavefunction.
With the isotope substitution, the first excited vibronic levels ($|j|=1$) \cite{Density} split 
and these eigenstates are approximately written as $\cos \phi$ and $\sin \phi$, respectively.
The density of the lower state $\cos \phi$ has maxima at $\phi = 0$ and $\pi$, 
which is the opposite to the correct one.

When the linear vibronic coupling is strong, the quadratic coupling term, 
$(W_E V_E^2/\omega_E^4) \cos 3\phi$, is also important \cite{Bersuker1989a, Bersuker2006a}.
Since the positions of the minima of the quadratic term is different from those induced by
the isotope effect, the distribution of the vibronic state depends 
on the strength of $V_E$.
When $W_E V_E^2/\omega_E^4 > \Delta_{\rm kin} + \Delta_{\rm pot}$, 
the distribution of the localization is characterized by the quadratic coupling term.
In the opposite case, $W_E V_E^2/\omega_E^4 < \Delta_{\rm kin} + \Delta_{\rm pot}$, 
the localization due to the isotope substitution is dominant.
Thus, the isotope effect is expected to be important in the region of the weak and 
intermediate vibronic coupling.

\section{``Orbital inversion'' symmetry}
Despite the warping of the trough, the tunneling splitting of 
the degenerate levels of $\hat{H}_\phi$ is not observed.
This degeneracy is not specific to the approximate Hamiltonian (\ref{Eq:H_rot})
but is a general property of the vibronic states of the 
vibronic Hamiltonian (\ref{Eq:HJT_iso}).
To prove the degeneracy, we introduce a unitary operator $\hat{\pi}$ defined by 
\begin{eqnarray}
 \hat{\pi} &=& \hat{\sigma}_y 
 \exp\left[ i \pi \left( \hat{N}_A + \hat{N}_\theta + \hat{N}_\epsilon \right)\right],
 \label{Eq:KJTISO}
\end{eqnarray}
where $\hat{N}_\mu$ is the number operator of the vibrational quanta
of the mode $\mu$, and $\hat{\sigma}_y$ the Pauli matrix. 
$\hat{\pi}$ is analogous to the operators $\mathcal{P}$ of Leung and Kleiner 
\cite{Leung1974a} and $\hat{P}$ of Bersuker and Polinger \cite{Bersuker1989a}
for the linear $E \otimes e$ JT problem.
$\hat{\pi}$ commutes with $\hat{\sigma}_y$ and 
anticommutes with $\hat{P}_\mu$, $Q_\mu$, 
$\hat{\sigma}_x$, and $\hat{\sigma}_z$, therefore, $[\hat{H}, \hat{\pi}] = 0$.
If $|\Psi \rangle$ is an eigenstate of $\hat{H}$, then
$|\hat{\pi}\Psi \rangle$ is also the eigenstate belonging to 
the same eigenenergy. 
$|\Psi\rangle$ and $|\hat{\pi}\Psi\rangle$ are different from 
each other because the irreducible representation of $\hat{\pi}$ is $b_1$.
If $|\Psi\rangle$ is $a_1$($b_1$), 
then $|\hat{\pi}\Psi\rangle$ is $b_1$($a_1$). 
Thereby, $|\Psi\rangle$ and $|\hat{\pi}\Psi\rangle$ are degenerate eigenstates
which are orthogonal to each other.
The eigenvalue of $\hat{\pi}$ is $\pm 1$ and the eigenstate is written as
$|\Psi_\pm\rangle = (|\Psi_{a_1}\rangle \pm i |\Psi_{b_1}\rangle)/\sqrt{2}$.
Here, $|\Psi_{a_1}\rangle$ and $|\Psi_{b_1}\rangle$ are degenerate 
vibronic states whose representations are $a_1$ and $b_1$, respectively.
Therefore, we can regard $\hat{\pi}$ as a parity operator in the space of the vibronic states,
resembling the time reversal operator acting on two wavefunctions of a Kramers doublet 
\cite{Abragam1970a}, in which spin should be replaced by orbital pseudospin $\tau=1/2$.

The two-fold degeneracy of each vibronic level 
of an isotope substituted JT system
is lifted in several cases.
First, the existence of the conical intersection is necessary for the degeneracy.
If we add $\hat{V} = \Delta \hat{\sigma}_z$ ($\Delta $ is real) 
to $\hat{H}$ to remove the conical intersection, 
the degeneracy of the vibronic state is lost because 
$[\hat{\pi}, {\hat{\sigma}}_z] \ne 0$ and the Hamiltonian does not have the 
parity symmetry.  
Second, quadratic (and even order)
JT coupling removes the degeneracy
of the vibronic level of isotope substituted system.
The quadratic coupling is not invariant under the 
operation $\hat{\pi}$ because $[\hat{\pi}, {Q}_\mu {Q}_\nu] = 0$, 
$[\hat{\pi}, {\hat{\sigma}}_x] \ne 0$, and
$[\hat{\pi}, {\hat{\sigma}}_z] \ne 0$.
The degeneracy of the vibronic level of a linear $E \otimes e$ JT system does not lift completely by the
quadratic coupling, while in the case of isotope substituted system the degeneracy is lost completely.
And finally, zero coupling to the totally symmetric mode in the Hamiltonian is relevant. 
If eq. (\ref{Eq:HEeJT}) has nonzero coupling to the $A$ mode, 
the Hamiltonian of the isotopomer has a term 
$\hat{V} = {V}_{A} {Q}_{A} {\hat{\sigma}}_0$
which does not commute with $\hat{\pi}$.
Here, ${V}_{A}$ is the vibronic coupling constant for the $A$ mode.

\section{Numerical Calculation}
\label{SEC:NumericalCalculation}
To confirm our analytical results on the isotope induced localization (eq. (\ref{Eq:psi2_rot})), we calculate the vibronic states 
numerically diagonalizing the vibronic Hamiltonian of X$_2$Y trinuclear system.
Owing to the reduction of the symmetry from $D_{3h}$ to $C_{2v}$, 
the $a_1'$ and $e'$ modes mix because
$a_1' \downarrow C_{2v} = a_1$ and $e' \downarrow C_{2v} = a_1 \oplus b_1$.
Consequently, the doubly degenerate electronic state couples
to the two $a_1$ modes and the $b_1$ mode.
\begin{eqnarray}
\hat{H} &=&
\sum_\alpha
\frac{1}{2}
\left(
 \hat{p}_\alpha^2 + \tilde{\omega}_\alpha^2 {q}_\alpha^2
\right)
\hat{\sigma}_0
\nonumber
\\
&+&
\sum_{i=1}^2 V_E C^\theta_{a_1(i)} {q}_{a_1(i)} \hat{\sigma}_z 
+
V_E C^\epsilon_{b_1} {q}_{b_1} \hat{\sigma}_x
.
\label{Eq:HEeJTISO}
\end{eqnarray}
Here, $\alpha = a_1(1), a_1(2), b_1$, 
and $\hat{p}_\alpha$ the conjugate momentum of $q_\alpha$.
The vibronic basis is a set of the products of the electronic states
$|\gamma\rangle$ and vibrational states 
$|n_{a_1(1)}, n_{a_1(2)}, n_{b_1}\rangle$,
$\{|\gamma\rangle |n_{a_1(1)}, n_{a_1(2)}, n_{b_1}\rangle
;\gamma = \theta, \epsilon, 0 \le n_{a_1(1)}+n_{a_1(2)}+n_{b_1} \le 30 \}$.
Parameters for the calculation are as follows:
$g_E = 3$, $\omega_E = 1$, $\omega_A = 1.3$, and $\delta = 1$.

\begin{figure}[tbh]
\onefigure[height=8.5cm]{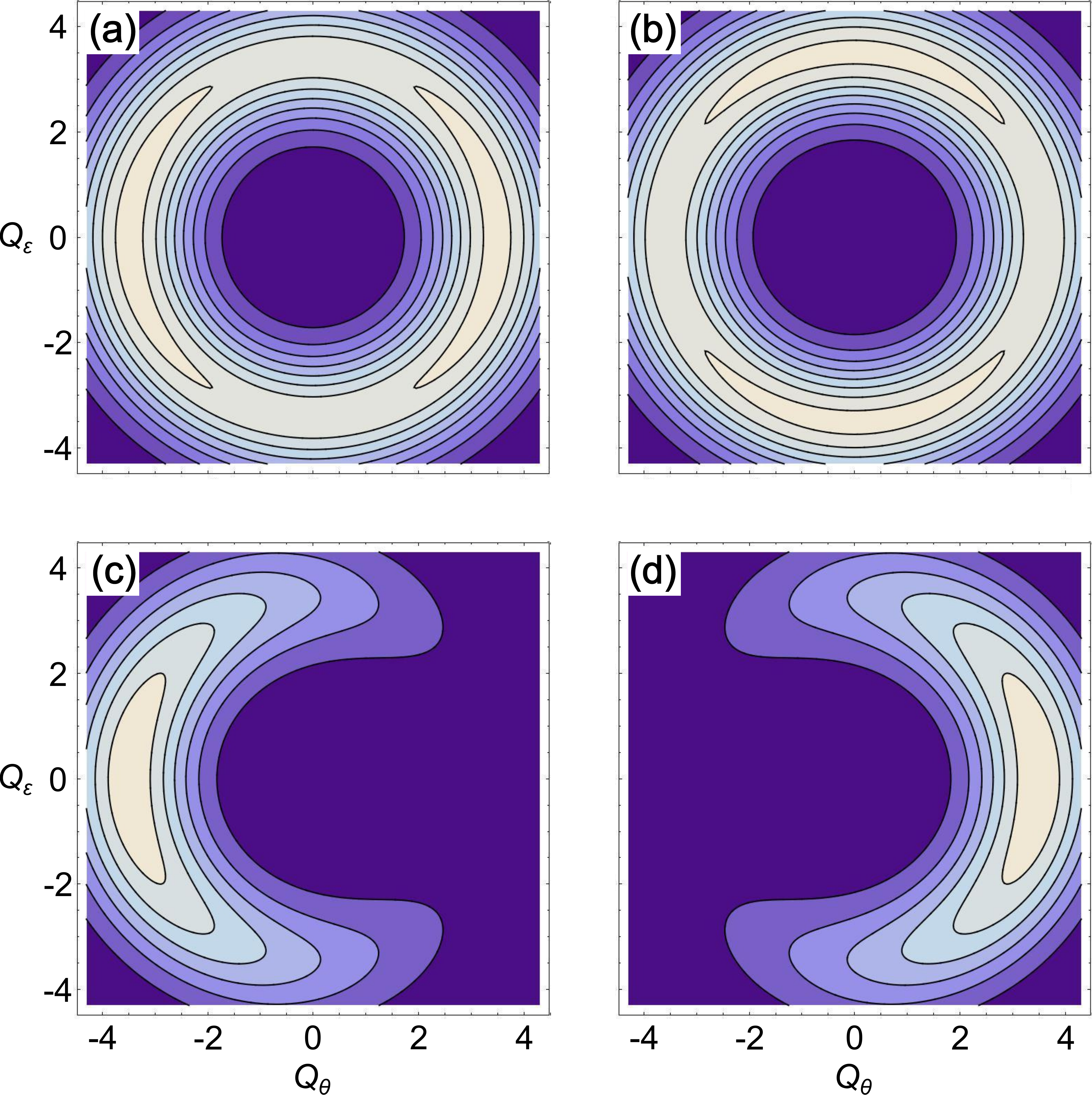}
\caption{
The densities of (a) one ground and (b) one first excited complex vibronic 
wavefunctions, and the (c) $a_1$ and (d) $b_1$ real vibronic wavefunctions in the ground state.
The calculations are done for $g_E = 3$, $\omega_E = 1$, $\omega_A = 1.3$, and $\delta = 1$.
The densities of the white areas are the largest and those of the dark blue the smallest.
The smallest value, the largest value, and the spacing on the contours of the plots (a) and (b) are 
0.002, 0.016, and 0.002, respectively. 
Those for (c) and (d) are 0.005, 0.030, and 0.005, respectively.
}
\label{Fig:Density}
\end{figure}

The densities $|\langle Q_\theta, Q_\epsilon|\Psi\rangle|^2$
of the ground and first excited vibronic states are shown in fig. \ref{Fig:Density}.  
The densities of the (a) ground and the (b) first excited states are localized around 
$\phi = 0, \pi$ and $\phi = \pm \pi/2$, respectively.
These distributions agree with the analytical results (\ref{Eq:psi2_rot}).
The densities of the (c) $a_1$ and (d) $b_1$ vibronic states 
are equivalent to each other after the rotation by $\pi$.
This clearly indicates that there is no tunneling splitting.

\section{Other Jahn--Teller system}
Similar isotope effects can be observed in other Jahn--Teller systems.
However, the effects in the triangular and other molecules should be
different from each other.
In the case of the triangular molecule, $\lambda$ in eq. (\ref{Eq:HJT_iso})
does not depend on the modes, while it does for other molecules.
For example, in the vibronic Hamiltonian of an octahedral molecule,
$\lambda$'s for the $\theta$ and $\epsilon$ modes are not the same.
In this case, this isotope effect is expected to be much stronger than that of X$_3$.
Moreover, the distribution of the density will depend on the sign of $\delta$.

In real molecules, the quadratic vibronic coupling constant $W_E$ is finite,
and the degeneracy of each level is lifted by isotope substitution.
As long as the vibronic coupling is weak or intermediate,
the splitting of the ground vibronic level due to the quadratic coupling
is expected to be smaller than the gap between the ground ($|j|=1/2$) and first 
excited ($|j|=3/2$) vibronic levels of the linear vibronic system.
Thus, the splitting of the ground levels could be experimentally observed with 
an appropriate method.
So far, the splitting of the ground vibronic levels of deuterated cyclopentadienyl 
radical has been reported \cite{Yu1993a,C5H5}.
Since the splitting cannot be seen as long as the Berry's phase is not taken into 
account, the observation of the splitting is a direct experimental evidence of the
existence of the Berry's phase in the isotopomers of this molecule.

\section{Conclusions}
We studied the isotope effect on the dynamical JT system.
The mechanisms of localization of JT deformations and vibronic wavefunctions
in isotope substituted systems are elucidated.
It is found that the localization in the trough is of potential type in the case of 
strong vibronic coupling, while the localization of the excited states becomes of 
kinetic type in the case of intermediate and weak coupling.
It is shown that the vibronic levels in the linear $E \otimes e$-problem remain double 
degenerate upon arbitrary isotope substitution because 
of an ``orbital inversion'' symmetry of the vibronic Hamiltonian.


\acknowledgments
N. I. and T. S. thank to hospitality of the quantum chemistry group 
in KU Leuven during their stay.
N. I. would like to acknowledge the financial support from Flemish Science Foundation (FWO).
T. S. thanks to Tatsuhisa Kato for useful discussion.
This work was supported in part
by the Japan Society for the Promotion of Science (JSPS)
through its Funding Program
for the Global COE Program
``International Center for Integrated Research
and Advanced Education in Materials Science'' (No. B-09)
of the Ministry of Education, Culture, Sports, Science and Technology (MEXT) of Japan.


\begin{thebibliography}{10}
\expandafter\ifx\csname url\endcsname\relax\def\url#1{\texttt{#1}}\fi

\bibitem{Bersuker1989a}
\Name{Bersuker I.~B. \and Polinger V.~Z.} \Book{Vibronic Interactions in
  Molecules and Crystals} (Springer--Verlag, Berlin and Heidelberg) 1989.

\bibitem{Bersuker2006a}
\Name{Bersuker I.~B.} \Book{The Jahn--Teller Effect} (Cambridge University
  Press, Cambridge) 2006.

\bibitem{Imada1998a}
\Name{Imada M., Fujimori A. \and Tokura Y.} \REVIEW{Revs. Mod.
  Phys.}{70}{1998}{1039}.

\bibitem{Tokura2000a}
\Name{Tokura Y. \and Nagaosa N.} \REVIEW{Science}{288}{2000}{462}.

\bibitem{Malguth2008a}
\Name{Malguth E., Hoffmann A. \and Phillips M.~R.} \REVIEW{Phys. Status Solidi
  (B)}{245}{2008}{455}.

\bibitem{Gunnarsson1997a}
\Name{Gunnarsson O.} \REVIEW{Revs. Mod. Phys.}{69}{1997}{575}.

\bibitem{Chibotaru2005a}
\Name{Chibotaru L.~F.} \REVIEW{Phys. Rev. Lett.}{94}{2005}{186405}.

\bibitem{Manini2010a}
\Name{Manini N. \and Tosatti E.} \Book{Jahn-Teller and Coulomb correlations in
  fullerene ions and compounds: From isolated ions to metal, insulator, and
  superconductor phases of alkali fulleride solids} (Lambert Acad. Publ.,
  Saarbrucken) 2010.

\bibitem{Lawler1964a}
\Name{Lawler R.~G., Bolton J.~R., Fraenkel J.~R. \and Brown T.~H.} \REVIEW{J.
  Am. Chem. Soc.}{86}{1964}{520}.

\bibitem{Carrington1965a}
\Name{Carrington A., Longuet-Higgins H.~C., Moss R.~E. \and Todd P.~F.}
  \REVIEW{Mol. Phys.}{9}{1965}{187}.

\bibitem{Yu1993a}
\Name{Yu L., Cullin D.~W., Williamson J.~M. \and Miller T.~A.} \REVIEW{J. Chem.
  Phys.}{98}{1993}{2682}.

\bibitem{Bruckmeier1994a}
\Name{Bruckmeier R., Wunderlich C. \and Figger H.} \REVIEW{Phys. Rev.
  Lett.}{72}{1994}{2550}.

\bibitem{Kramer1999a}
\Name{Kr{\"{a}}mer H.-G., Keil M., Suarez C.~B., Demtr{\"{o}}der W. \and Meyer
  W.} \REVIEW{Chem. Phys. Lett.}{299}{1999}{212}.

\bibitem{Worner2007a}
\Name{{W\"{o}rner} H.~J. \and Merkt F.} \REVIEW{J. Chem.
  Phys.}{126}{2007}{154304}.

\bibitem{Melnik2011a}
\Name{Melnik D.~G., Liu J., Chen M.-W., Miller T.~A. \and Curf R.~F.}
  \REVIEW{J. Chem. Phys.}{135}{2011}{094310}.

\bibitem{Ham1987a}
\Name{Ham F.~S.} \REVIEW{Phys. Rev. Lett.}{58}{1987}{725}.

\bibitem{Density}
In this case Eq. (\ref{Eq:psi2_rot}) does not apply because the matrix elements
  of the terms proportional to $\cos 2\phi$ are nonzero between these
  degenerate vibronic states, and one should apply the degenerate perturbation
  theory.

\bibitem{Leung1974a}
\Name{Leung C.~H. \and Kleiner W.~H.} \REVIEW{Phys. Rev. B}{10}{1974}{4434}.

\bibitem{Abragam1970a}
\Name{Abragam A. \and Bleaney B.} \Book{Electron Paramagnetic Resonance of
  Transition Ions} (Claredon Press, Oxford) 1970.

\bibitem{C5H5}
  In the case of the asymmetrically deuterated cyclopentadienyl
  radical, the splitting of the ground vibronic level is due to the fourth
  order vibronic coupling. The proof of the splitting is almost the same as
  that by the quadratic coupling.

\end{thebibliography}

\end{document}